\documentclass[aps,prl,10pt,nofootinbib,sectionnumbers,reprint,longbibliography]{revtex4-2}

\usepackage[utf8]{inputenc}
\usepackage{amsmath,amssymb}
\usepackage{graphicx}
\usepackage{textcomp}
\usepackage[
  colorlinks=true,
  linkcolor=blue,      
  citecolor=green,     
  urlcolor=magenta     
]{hyperref}
\usepackage{xcolor}
\usepackage{physics}
\usepackage{lipsum}
\usepackage{graphicx} 

\def\IB#1{\boldsymbol{#1}} 
\def\W/!i#1{\Wi} 

\usepackage{accents}
\usepackage{bm}

\makeatletter
\renewcommand{\@makefnmark}{%
  \hbox{\textcolor{olive}{\@thefnmark}}}
\makeatother


\usepackage[font=small,labelfont=bf,justification=justified,singlelinecheck=false]{caption}
\providecommand{\vs}[1]{}
\renewcommand{\vs}[1]{\textcolor{red}{VS: \textit{#1}}}

\providecommand{\ac}[1]{}
\renewcommand{\ac}[1]{\textcolor{blue}{AC: \textit{#1}}}

\begin{document}

\title{Propulsion and far-field hydrodynamics of linked-sphere microswimmers with viscoelastic deformability}

\author{Vimal Singh}
\author{Akash Choudhary$^\dagger$}
\footnotetext[2]{Corresponding author: achoudhary@iitk.ac.in}
\affiliation{Department of Chemical Engineering, Indian Institute of Technology Kanpur, India}

\begin{abstract}
Viscoelasticity governs the locomotion strategies of deformable microorganisms, rendering it a fundamental mechanical property of microbial motility and an integral component in the design of envisioned microbots.
Recent studies have shown that it can enable effective propulsion through non-reciprocal body deformations, even under time-reversible actuation.
In this work, we investigate the dynamics of model microswimmers driven by reciprocal actuation, wherein the passive body exhibits viscoelastic deformability. 
We consider two linked-sphere designs, distinguished by the location of actuation: applied at one end (3-sphere design) or at the midpoint of the swimmer body (4-sphere design).
Adopting Kelvin-Voigt deformability, we characterize the kinematic performance of both designs: the three-sphere swimmer possesses an optimal actuation frequency, while the four-sphere swimmer exhibits a critical frequency at which the locomotion direction reverses.
We examine the swimmer’s far-field hydrodynamic signature and find that resulting flow field is characterized by dominant dipolar and quadrupolar contributions, whose magnitudes are sensitive to the relative length of the actuator segment. 
\end{abstract}

\maketitle

{\section{Introduction}}
\vspace{-0.5cm}
Self-propulsion of microscopic organisms and bots is governed by low-Reynolds-number hydrodynamics, a regime where viscous effects completely dominate over inertial force \cite{purcell1977life}.
In this Stokes regime, Purcell’s scallop theorem asserts that a microscopic swimmer executing a reciprocal sequence of body deformations cannot achieve any net displacement over a complete cycle.
\citet{najafi2004simple} showed that a minimal three-sphere microswimmer requires at least two degrees of freedom to break symmetry over one cycle of deformation. They demonstrated that $O(\epsilon)$ deformations yield $O(\epsilon^2)$ locomotion due to inter-sphere hydrodynamic interactions. Later, \citet{golestanian2008analytic} detailed this framework for harmonic, asymmetric, and noisy deformations, finding that a $\pi/2$ phase difference between the deformations of the two arms is optimal. Recently, \citet{nasouri2019efficiency} used boundary element simulations to explore efficiency limits and optimal strokes across a wide variety of actuation and geometric parameters.
While early studies of linked-sphere microswimmers focused primarily on their locomotion and kinematic performance, \citet{pooley2007hydrodynamic} and \citet{alexander2009hydrodynamics} turned attention to the hydrodynamic signatures these swimmers impart on the surrounding fluid.
A recent review by \citet{yasuda2023generalized} encapsulates further advancements by categorizing diverse extensions to the three-sphere model, including elastic, stochastic, and autonomous swimmers.

The restriction imposed by the scallop theorem can be circumvented either by exploiting the nonlinearity of the surrounding medium or by employing swimmer geometries whose passive response inherently breaks the time-reversal symmetry of the imposed actuation \cite{lauga2011life}.
The former has been explored extensively by considering fluid regimes that exhibit inertial \cite{klotsa2015propulsion} or viscoelastic \cite{datt2018two,yasuda2020reciprocal,eberhard2023reciprocal} or other rheological effects \cite{yasuda2018three}. Among the latter, geometric approaches such as the push-me-pull-you swimmer exploit volume exchange between linked bodies to generate non-reciprocal strokes, drawing inspiration from euglenoid movements \cite{avron2005pushmepullyou,silverberg2020realization}. A complementary route exploits the passive elastic deformability of the swimmer body itself: \citet{nasouri2017elastic} demonstrated that a two-sphere swimmer with one neo-Hookean elastic sphere can achieve net locomotion under purely reciprocal actuation, as the front-back asymmetric deformation of the elastic body introduces the required symmetry-breaking.
In a parallel development within the three-sphere framework, \citet{montino2015three} investigated an elastic three-sphere microswimmer in which one arm is a passive elastic spring while the other undergoes prescribed periodic actuation. 
They demonstrated that a single actuated degree of freedom suffices for locomotion, as the competition between elastic restoring forces and viscous drag induces a phase lag in the passive arm response, generating non-reciprocal deformations. 
Subsequent work introduced active coordination via muscle-like contraction models \cite{montino2017dynamics}; however, in both formulations the swimming direction remains structurally constrained. 
While recent work has demonstrated the scope for bi-directionality through multi-link elastic architectures \cite{alouges2023limiting,levillain2025bi}, the role of viscoelastic passive deformability---wherein the interplay between elastic and viscous timescales of the body introduces a frequency-dependent optimality criterion---has not been explored.
Furthermore, the far-field hydrodynamic signature of such deformable microswimmers remains entirely uncharacterized, despite its central role in determining inter-swimmer interactions and collective dynamics.

Viscoelasticity is a defining mechanical characteristic of several microorganisms, with significant consequences for their locomotion strategies \cite{spagnolie2010optimal,espinosa2013fluid,thomases2017role}. 
From an engineering standpoint of microbot design, the Kelvin–Voigt model represents the simplest yet effective constitutive framework for capturing the viscoelastic flexural response of soft polymeric flagella \cite{liu2024propulsion}.
The present study examines the effect of this simplest viscoelastic deformability on propulsion dynamics and far-field signature of a linked-sphere minimal microswimmer. We consider both asymmetric and symmetric actuating problems by modeling 3-sphere and 4-sphere systems, respectively (see Fig. \ref{fig:schematic1},\ref{fig:schematic2}).
The passive links of the microswimmer are modeled using Kelvin-Voigt elements, characterized by an elastic modulus and a viscous damping coefficient to capture their internal mechanical response. 
Using numerical simulations (in \S\ref{sec:2}) and perturbation theory (in \S\ref{sec:3}), we identify an optimal actuation frequency at which propulsion is maximized for 3-sphere swimmer, governed by the balance between viscoelastic relaxation and the degree of kinematic non-reciprocity; for the four-sphere swimmer, we demonstrate the existence of a critical frequency marking a reversal in the direction of locomotion.
We further (in \S\ref{sec:4}) derive far-field dipolar and quadrupolar hydrodynamic signatures using multipole expansion framework \cite{alexander2009hydrodynamics}, which govern long-range interactions with boundaries and other swimmers.

\vspace{0.5cm}

\section{Model and Numerical Results}\label{sec:2}
\begin{figure}[t]
     \centering
     \includegraphics[width=0.93\linewidth]{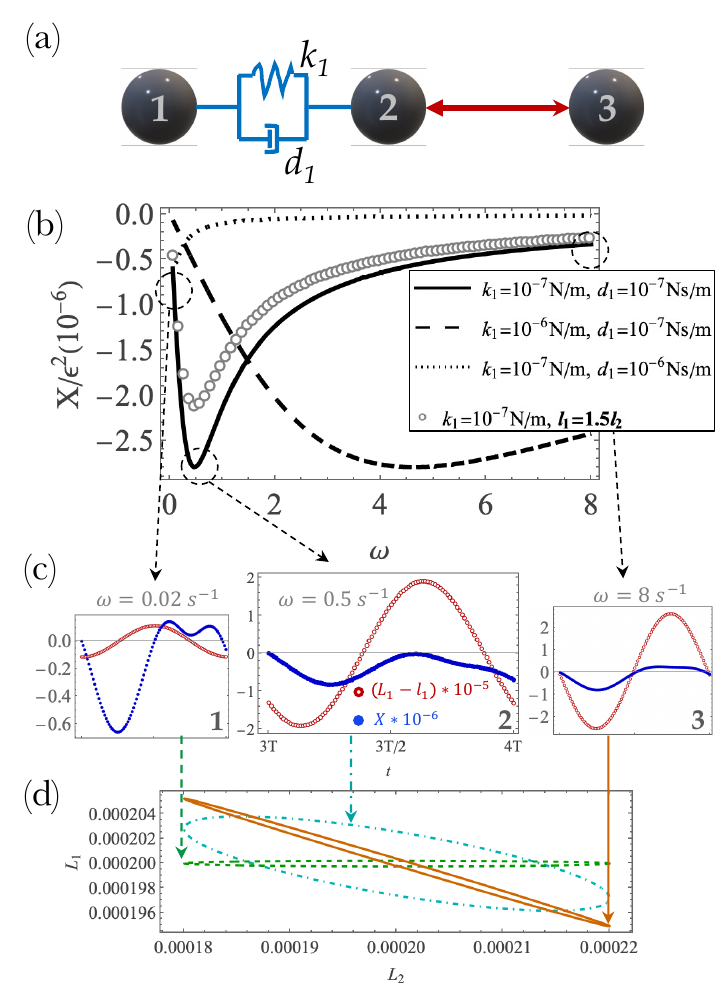}
     \caption{
     \footnotesize
     {(a)} Schematic showing the 3-sphere swimmer. The red link represents the actuation and the blue link denotes that passive part of swimmer's body is undergoing viscoelastic deformation modeled using the Kelvin-Voigt model. 
     Here $k$ represents the elastic modulus and $d$ is the damping coefficient. 
    {(b)} Displacement per unit cycle for various frequencies and three combination of viscoelastic properties for $l_1=l_2=2*10^{-4} m,$ $a=0.1*10^{-4} m,$ $ \mu=8.9*10^{-4}$ Pa.s. Circles represent the plot for a swimmer that is geometrically asymmetric $l_1=2.4*10^{-4}$ and $ l_2=1.6*10^{-4} m $. 
     {(c)} Temporal change in deviation from natural length ($L_1-l_1$) and net displacement of the swimmer $(X)$ over a time period $T=2\pi/\omega$ away from $t=0$ such that initial transients no longer contribute. 
    {(d)}  Phase portraits representing the temporal evolution $L_1$ and $L_2$ for different frequencies.
     }
     \label{fig:schematic1}
 \end{figure}

\begin{figure}[t]
     \centering
     \includegraphics[width=1.03\linewidth]{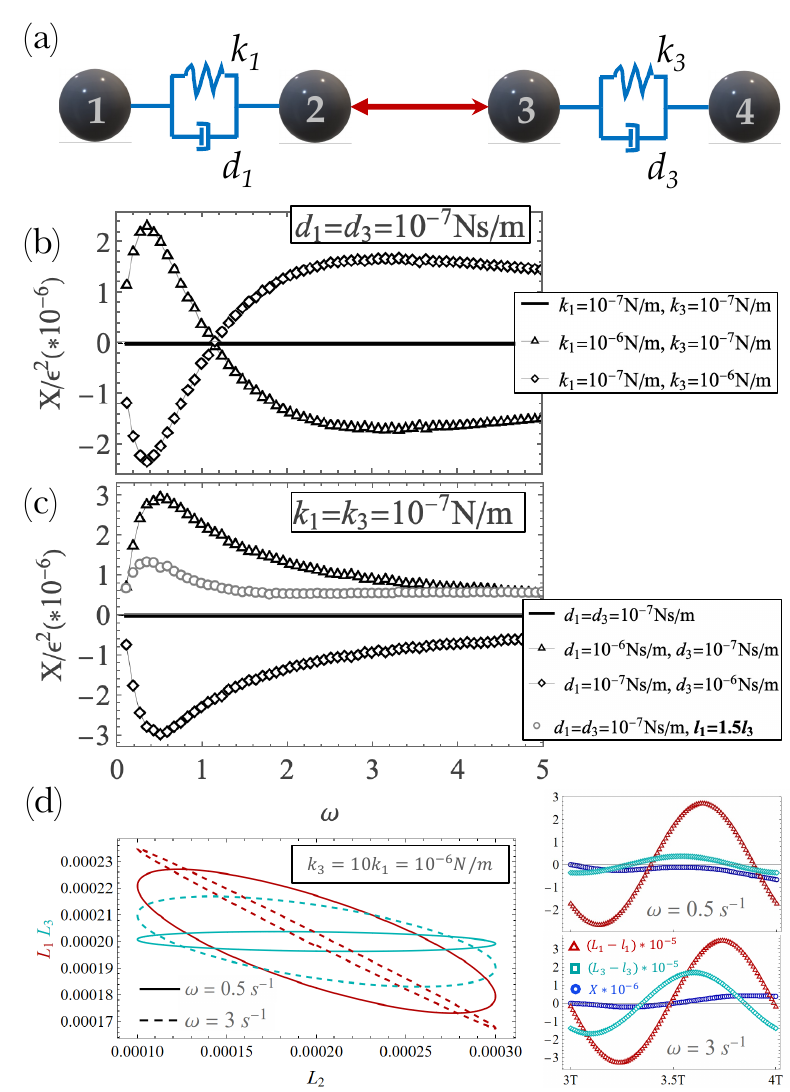}
     \caption{
     \footnotesize
   {(a)} Schematic showing the 4-sphere swimmer with actuating link in the middle. 
     {(b,c)} Displacement per unit cycle for various frequencies and various combinations of viscoelastic properties for $l_1=l_2=2*10^{-4} m,$ $a=0.1*10^{-4} m,$ $ \mu=8.9*10^{-4}$ Pa.s.  Open circles represent geometrically asymmetric swimmer.
    {(d)} Temporal evolution of the two shape parameters $L_1-L_2$ (Red) and $L_3-L_2$ (Cyan) for high and low frequencies. 
     The adjacent insets show temporal change in deviation from natural lengths and the net displacement of the swimmer $(X)$ over a cycle of deformation.   
     }
     \label{fig:schematic2}
 \end{figure}

We consider two swimmer models within a single framework. Figures \ref{fig:schematic1}(a) and \ref{fig:schematic2}(a) illustrate linked-sphere microswimmers capable of self-propulsion in a Newtonian fluid through the periodic actuation of a subset of their components.
The prescribed actuating length of the link is $L_2(t)$, whereas length of the passive viscoelastic link is $L_1(t)$ and $L_3(t)$ for three- and four-sphere swimmer, respectively.
The deformation-induced displacement of spheres (of identical size $a$) here is along the length axis.
The activity is imparted by the active link as: $L_2(t)=l_2 (1+ \epsilon \sin \, \omega t)$, where $l_2$ is the natural or equilibrium length of the passive link and $\epsilon$ represents the amplitude of deviation due to actuation with frequency $\omega$.
The low Reynolds hydrodynamics around the swimmer obey the Stokes equations for the velocity and pressure field ($\mu \nabla^2 \IB{v}  = \nabla p$), along with the incompressibility condition.
Assuming that the spheres are sufficiently far apart ($a \ll L_i$) at all times, their motion with unidirectional velocity $V_i$ predominantly generates a Stokeslet that interacts with the other spheres. This effectively modifies the mobility factors from the Stokes drag and can be accounted using the following linear relationship: 
\begin{equation}
V_i = \sum_{j=1}^{n} {\mathcal{M}}_{ij}  f_j, \quad
\mathcal{M}_{ij} = \frac{1}{6 \pi \mu a} \left\lbrace \delta_{ij} + \frac{3}{2} \left[ \frac{a}{\chi_{ij}(t)} \right]_{i \neq j} \right\rbrace,
 \label{eq:mob}
\end{equation}
where \(\mathcal{M}_{ij}\) is the modified mobility tensor accounting for the far-field Stokeslet interactions \cite{batchelor1976brownian}, \(f_i\) is the force exerted by the sphere on the fluid and $\chi_{ij}$ is the dynamic distance between $i^{\text{th}}$ and $j^{\text{th}}$ sphere. For example, $\chi_{12} = \chi_{21} = L_1$, $\chi_{13} = \chi_{31} = L_1 + L_2$, and $\chi_{14} = \chi_{41} = L_1 + L_2 + L_3$.

The expressions for velocity in Eq. (\ref{eq:mob}) can be obtained by employing the additional equations from force and kinematic balance, as described below.
First, the swimmer is freely suspended and thus satisfies the force-free condition $\sum_{i}^{n}f_i =0$. 
Second, the free sphere at the ends that is connected only to the passive viscoelastic arm experiences the following force balance due to actuation-induced motion as:
\begin{equation}\label{eq:fb}
\begin{split}
    k_1*(L_1-l_1)+d_1*\dot{L}_1=f_1
    \\
     k_3*(L_3-l_3)+d_3*\dot{L}_3=-f_4
\end{split}
\end{equation}
Here $k_i$ and $d_i$ represent the elastic modulus and damping coefficients, and $l_i$ denote the rest length of the viscoelastic links.
For the 3-sphere swimmer, only the first relation is used, whereas for the 4-sphere swimmer, both relations are used.
Finally, the following kinematic relations are implied by the geometry:
\begin{equation}\label{eq:kin}
    \dot{L}_i = V_{i+1} - V_i.
\end{equation}
The aforementioned equations are solved to obtain equations of $\dot{L}_1$ and $\dot{L}_3$. These  can be solved numerically with initial conditions of natural length (i.e., $L_i(t=0)=l_i$), which finally yield the estimates of velocity $V_i$. The average velocity for 3-sphere swimmer $\overline{V} = \Sigma_i V_i/N$ is straightforwardly evaluated.
Finally, the displacement can be obtained as $ X(t') = \int_{0}^{t'} \overline{V}(t) dt$.
\\


\textbf{For the 3-sphere swimmer}, Fig.\ref{fig:schematic1} (b) shows the displacement per actuation cycle across a range of frequencies. Two key observations emerge: (\textit{i}) the swimmer invariably migrates in the direction of the passive viscoelastic link (in the negative direction for this specific geometry), and (\textit{ii}) there exists an optimal \textit{resonant} frequency for each combination of material viscoelasticity ($k$ and $d$) and fluid viscosity. We address each in turn.

The first observation stems from the asymmetry in the hydrodynamic interactions encoded in $\mathcal{M}_{ij}$, specifically via $\chi_{12}$ distance.
During the extension phases of $L_2$, the passive link $L_1$ undergoes contraction. As a result, spheres 1 and 2 move closer to each other during the backward stroke, leading to stronger hydrodynamic interactions and hence a larger backward displacement of the swimmer.
This action is followed by $L_2$'s contraction, which expands the passive $L_1$ with overall weaker hydrodynamic interaction between spheres 1 and 2, producing a smaller forward displacement.\footnote[2]{This asymmetry between strong and weak hydrodynamic coupling within the actuation cycle is analogous to the mechanism underlying propulsion in the Najafi–Golestanian swimmer \cite{najafi2004simple}. In that system, although both links are non-reciprocally actuated, the net swimming direction is governed by the actuation in the contracted state when spheres are closest.}
This is corroborated by temporal plot in Fig. \ref{fig:schematic1}(c,2), which shows that the displacement ($X$) correlates with $L_1 - l_1$ across various frequencies and material and fluid properties: when $L_1 - l_1$ is negative (contracted $L_1$), the magnitude of the negative displacement is larger than that of the positive displacement in the expanded state.


We next examine the second observation: how actuation frequency governs displacement magnitude and its optimality. Figs.~\ref{fig:schematic1}(c,1-3) illustrate the deformation-displacement dynamics for a few frequencies, revealing that the observed optimality is governed by two competing factors.
In the low-frequency limit, the passive viscoelastic link responds quasi-statically, following the $L_2$'s actuation with near-quadrature ($\pi/2$) phase lag, i.e., $L_1 - l_1 \sim -\cos\omega t$.
However, the deformation magnitude of $L_1$ is negligible for such slow actuations, yielding a correspondingly negligible net displacement per actuation cycle.
At the other extreme of very high frequencies, the dashpot of the passive link dominates, enforcing an anti-phase deformation of $L_1$ (i.e., $L_1-l_1 \sim  - \sin \omega t$), such that $L_2$ expands and $L_1$ contracts, both sinusoidally. 
In this regime, despite the significant deformation amplitude, the deformation is nearly symmetric and lacks the phase asymmetry necessary for non-reciprocity.
Hence, the optimal locomotion is found at intermediate frequencies, governed by a trade-off between deformation magnitude and the degree of asymmetric phase lag between $L_1$ and $L_2$.
Fig. \ref{fig:schematic1}(d) corroborates this picture, showing that the interplay of these two factors yields an optimal deformation state, quantified by the enclosed area of the $L_1 - L_2$ phase portrait (representing magnitude of displacement per unit cycle).

We note that increasing the elastic modulus of the passive link ($k$) reduces its characteristic relaxation time, enabling a more rapid response to the imposed actuation and thus diminishing $L_1$ deformation at a given frequency. 
Consequently, the optimal frequency shifts to higher values, as shown in Fig. \ref{fig:schematic1}(b). Furthermore, increasing viscous damping attenuates the peak deformation of $L_1$ and correspondingly suppresses net locomotion. 
Finally, increasing the natural length of the passive link ($l_1$) reduces the relative deformation magnitude, yielding a smaller net displacement per actuation cycle.
\\


\textbf{{For the 4-sphere swimmer}}, if the geometric and viscoelastic properties are symmetric across the actuating link ($k_1=k_2, \; d_1=d_2, \; l_1=l_2$), there can be no locomotion due to complete in-phase deformation between $L_1 \, \& \, L_3$;
this is shown by the zero-displacement line in Fig.\ref{fig:schematic2} (b).
When the viscoelastic properties are asymmetric---for example, $k_3 > k_1$, as shown in Fig. \ref{fig:schematic2}(b)---net locomotion emerges whose direction is, interestingly, switchable with the actuation frequency.
For low frequencies, the stiffer viscoelastic link $L_3$ deforms relatively less, effectively resembling a 3-sphere swimmer with a single deforming link $L_1$. Hence, the swimmer with $k_1 < k_3$ moves backwards towards $L_1$. 
This behavior is illustrated in Fig. \ref{fig:schematic2}(d): the solid lines show that area under the $L_1-L_2$ portrait is larger than $L_3-L_2$.
At higher frequencies, conversely, the deformation in $L_3$ dominates, which reverses the direction of net displacement.
The adjacent plots of the relative deformations and the sign of $X$ further corroborate this frequency-dependent direction of displacement.
For the opposite stiffness contrast, $k_3 < k_1$, these trends are reversed.
Furthermore, Fig.~\ref{fig:schematic2}(c) shows that asymmetry in viscous damping can also determine the direction of locomotion. For instance, for $d_3 > d_1$, the deformation of $L_3$ is suppressed across all frequencies and thus the swimming is towards $L_1$.
Finally, it is interesting to note that asymmetry in geometry can also facilitate locomotion even when the viscoelastic properties are symmetric. As shown by the open circles in Fig.~\ref{fig:schematic2}(c), a larger natural length $l_1$ effectively mimics a situation in which the deformations of $L_1$ are damped and locomotion is towards $L_3$.

While these numerical results reveal a rich dependence on actuation frequency, geometric and viscoelastic asymmetries, their combined effects are not easily disentangled. 
To rationalize these trends, we pursue analytical progress in the small-amplitude limit, introducing in the following section a non-dimensional formulation and a perturbation expansion in the actuation amplitude.


\section{Analytical insights under weak deformation regime}\label{sec:3}

In this section, we derive the asymptotic solutions to three and four-sphere swimmers. First, we describe the non-dimensional variables as following:
\begin{equation}
    \Lambda_i = L_i/l, \; \tau = \omega t, \; \alpha=a/l, \; \ell_i = l_i/l, \; v_i=V_i/\omega l,
\end{equation}
where $l=l_1+l_2+l_3$ is the total natural length of the swimmer.
We introduce the dimensionless elasticity number (ratio of actuation time scale to elastic relaxation time scale) $ \mathcal{K}_i={k_i}/{(\omega \mu l)}$
and damping number (ratio of internal to external viscous damping) $ \mathcal{D}_i={d_i}/{(\mu l)}$.
\\

\textbf{For a 3-sphere swimmer}, we  denote $ \mathcal{K}_1=\mathcal{K} $ \& $ \mathcal{D}_1=\mathcal{D} $ and obtain the following evolution equation for viscoelastic link:
\begin{equation}\label{eq:3S_Lambda}
   \dot{\Lambda}_1 = \frac{-\mathcal{K} \mathbb{G}}{1+\mathcal{D} \mathbb{G}} (\Lambda_1 - \ell_1) -\frac{ \mathbb{F}} {1+\mathcal{D} \mathbb{G}} \dot{\Lambda}_2,
\end{equation}
where $\mathbb{F}(\Lambda_i,\alpha)$ and $\mathbb{G}(\Lambda_i,\alpha)$ are the following dimensionless geometric functions:
\begin{align}\label{eq:FG}
    \mathbb{F} &=  
\frac{-2 \Lambda_1 \Lambda_2 (\Lambda_1 + \Lambda_2)
+ 3 \alpha (\Lambda_1^2 + \Lambda_1 \Lambda_2 + \Lambda_2^2)}
{2 \Lambda_1 (3 \alpha - 2 \Lambda_2) (\Lambda_1 + \Lambda_2)}, 
\nonumber
\\
\mathbb{G} &= 
\frac{
{\alpha\Lambda_2}
\left(
-\frac{2}{\alpha}
+ \frac{3}{\Lambda_1}
+ \frac{3\Lambda_1}{\Lambda_1\Lambda_2 + \Lambda_2^2}
\right)^2
}{
24(3 \pi \alpha - 2 \pi \Lambda_2)
}
+
\frac{1}{3 \pi \alpha}
- \frac{1}{2 \pi \Lambda_1}.
\end{align}
We now perform perturbation expansion of the viscoelastic link, assuming weak deformations ($\epsilon \ll 1$).
\begin{equation}\label{eq:pert}
    \Lambda_1 (\tau)=\lambda^{(0)} + \epsilon \lambda^{(1)} (\tau) + O(\epsilon^2), \text{ where } \lambda^{(0)}=\ell_1.
\end{equation}
Here $\lambda^{(1)}$ captures the first order temporal deformation, and prescribe the actuating link as $\Lambda_2 = \ell_2(1+\epsilon \sin \tau)$.
We now substitute this in Eq.(\ref{eq:3S_Lambda}) and extract the first order terms as:
\begin{equation}\label{eq:3S_lambda}
    \dot{\lambda}^{(1)}= \frac{-\mathcal{K} \mathbb{G}^{(0)} \lambda^{(1)} -\mathbb{F}^{(0)} \ell_2 \cos\tau}{1+ \mathbb{G}^{(0)} \mathcal{D}},
\end{equation}
where $\mathbb{G}^{(0)} (\ell_i,\alpha)$ and $\mathbb{F}^{(0)}(\ell_i,\alpha)$ are zeroth order versions of Eq.(\ref{eq:FG}).
The solution for Eq.(\ref{eq:3S_lambda}) is
  \begin{equation}\label{eq:3S_lambda-sol}
     \lambda^{(1)}=-A\, \ell_2 \sin{\tau} + B\, \ell_2 \left[\exp(\frac{-\mathbb{G}^{(0)} \mathcal{K} \tau}{\mathbb{G}^{(0)} \mathcal{D}+1}) - \cos{\tau}\right].
 \end{equation}
Here the exponential term decays rapidly and does not contribute from second cycle onward. 
We also note that the coefficients $A$ and $B$ multiplying in-phase and out-of-phase contributions, respectively, are dependent on elasticity and damping number as:
 \begin{align}
 A(\mathcal{K},\mathcal{D})&=\frac{\mathbb{F}^{(0)} (\mathbb{G}^{(0)} \mathcal{D}+1)}{\mathbb{G}^{(0)\,2} \mathcal{K}^2+(\mathbb{G}^{(0)} \mathcal{D}+1)^2},
\nonumber
\\
B(\mathcal{K},\mathcal{D})&=\frac{\mathbb{F}^{(0)} \mathbb{G}^{(0)} \mathcal{K}}{\mathbb{G}^{(0)\,2} \mathcal{K}^2+(\mathbb{G}^{(0)} \mathcal{D}+1)^2}.
 \end{align}
Interestingly, the out-of-phase contribution in Eq.(\ref{eq:3S_lambda-sol}) corresponds to non-reciprocal deformation and is solely responsible for locomotion (as also subsequently shown explicitly in eq.\ref{eq:3S_X}).
In the limit of no damping ($\mathcal{D} \to 0$), these results reproduce the expressions for purely elastic swimmer \cite{montino2015three}.
At high actuation frequencies, the elastic relaxation time becomes comparatively large ($\mathcal{K} \to 0$), causing the out-of-phase coefficient $B$ to vanish, and consequently locomotion ceases.

\begin{figure}[h]
     \centering
     \includegraphics[width=0.8\linewidth]{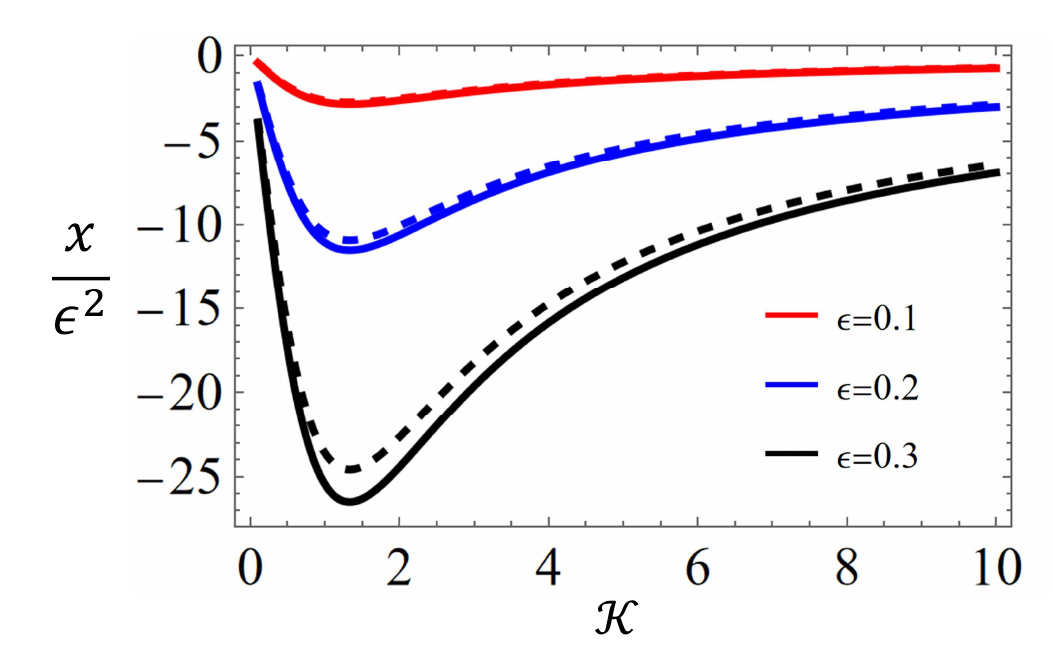}
     \caption{
     \footnotesize 
      Comparison between numerical simulation (solid line) and Eq. (\ref{eq:3S_X}) (dashed line) for displacement per unit cycle for various $\mathcal{K}$ and three combination of $\epsilon$  for $\mathcal{D}=1$, $\ell_1=\ell_2=0.5$, $\alpha =0.1/4$.}
     \label{fig:schematic3}
 \end{figure}

To obtain the leading order expression for mean velocity, we obtain the dimensionless version of velocity equation that can be obtained from rearranging Eqs.(\ref{eq:mob}-\ref{eq:kin}):
\begin{equation}\label{eq:Vbar}
  \overline{v}=\frac{\alpha}{6}\left[\left(\frac{\dot{\Lambda_2}-\dot{\Lambda_1}}{\Lambda_1+\Lambda_2}
  +
  \frac{2\dot{\Lambda_1}}{\Lambda_2}-\frac{2\dot{\Lambda_2}}{\Lambda_1}\right)
  +
  \left(\frac{\dot{\Lambda_2}}{\Lambda_2}-\frac{\dot{\Lambda_1}}{\Lambda_1}\right)\right],  
\end{equation}
where the second term averages to zero over each deformation cycle  and is thus neglected in subsequent analysis \cite{golestanian2008analytic}. 
It is worth noting that the above expression for a generic 3-sphere swimmer, consistent with \citet{golestanian2008analytic}, neglects the minute $O(\alpha^2)$ terms; the full expression is, however, retained in the numerical calculations. 
Expanding $\Lambda_1$ and $\Lambda_2$ according to Eq.(\ref{eq:pert}), we obtain the leading order contribution as:
\begin{align}\label{eq:3S_v}
    \overline{v} =  
O(\epsilon)
+   & \epsilon^2  \frac{\alpha}{6} \left( 
- \frac{(\lambda^{(1)} + \ell_2 \sin \tau)\,(\ell_2 \cos \tau - \dot{\lambda}^{(1)})}{(\ell_1 + \ell_2)^{2}} +
\right.
\nonumber
\\
& \qquad \;\; \left.
\frac{2 \lambda^{(1)} \ell_2 \cos \tau}{\ell_1^{2}}
- \frac{2 \sin \tau\, \dot{\lambda}^{(1)}}{\ell_2}
\right) .
\end{align}
Here, the first order term does not contribute towards time-averaged displacement.
Substituting Eq.(\ref{eq:3S_lambda},\ref{eq:3S_lambda-sol}) and neglecting the short-lived transients, we obtain the time-averaged displacement, $ x = X/l = \int_{0}^{2\pi} \overline{v}(\tau) d\tau$, as:
\begin{equation}\label{eq:3S_X}
x = \frac{-\pi \alpha B}{3} \epsilon^2 \left(  
\frac{
\ell_1^{4}
+ 2 \ell_1^{3} \ell_2
+ \ell_1^{2} \ell_2^{2}
+ 2 \ell_1 \ell_2^{3}
+ \ell_2^{4}
}{
\ell_1^{2} (\ell_1 + \ell_2)^{2}
}
\right).
\end{equation}
We note that the sign of net displacement per unit cycle is negative i.e., towards the passive viscoelastic link and the leading order contribution is $O(\epsilon^2 \alpha)$. This direction is independent of geometric anisotropy.
The proportionality with $B$ indicates that viscous damping reduces the net displacement.
Fig. \ref{fig:schematic3} plots average displacement expression Eq.(\ref{eq:3S_X}) shows a close agreement with numerical for small $\epsilon$ values. 
\\

\textbf{For a 4-sphere swimmer}, we obtain the following evolution equations for viscoelastic links in non-dimensional form:
\begin{align}\label{eq:4s_Lambda}
  \dot{\Lambda}_1=-\frac{\mathcal{K}_1\mathbb{G}_1}{1+\mathcal{D}_1\mathbb{G}_1 }(\Lambda_1 - \ell_1)+\frac{\mathbb{F}_1\, \dot{\Lambda}_2}{1+\mathcal{D}_1\mathbb{G}_1}-\frac{\mathbb{H}_1\, \dot{\Lambda}_3}{1+\mathcal{D}_1\mathbb{G}_1}, 
\nonumber
\\
 \dot{\Lambda}_3=-\frac{\mathcal{K}_3\mathbb{G}_3}{1+\mathcal{D}_3\mathbb{G}_3 }(\Lambda_3 - \ell_3)+\frac{\mathbb{F}_3 \,  \dot{\Lambda}_2}{1+\mathcal{D}_3\mathbb{G}_3} -\frac{\mathbb{H}_3\, \dot{\Lambda}_1}{1+\mathcal{D}_3\mathbb{G}_3}.
\end{align}
Unlike three-sphere swimmer, the deformation of each viscoelastic link is now influenced not only by the actuation of the central link $\Lambda_2$ but also by the deformation rate of the other passive link, leading to hydrodynamic coupling terms proportional to $\mathbb{H}_i$, $\mathbb{F}_i$ and $\mathbb{G}_i$. 
Assuming weak deformations, we perform perturbation expansion in $\epsilon$ for $\Lambda_1$ and $\Lambda_3$: $ \Lambda_i (\tau)=\lambda^{(0)}_i + \epsilon \lambda^{(1)}_i (\tau) + O(\epsilon^2)$, where $\lambda^{(0)}_1=\ell_1, \lambda^{(0)}_3=\ell_3$ and $\lambda^{(1)}_1$ and $\lambda^{(1)}_3$ capture the first-order temporal deformations. 
In similar notation, the actuating link is prescribed as $\Lambda_2=\ell_2(1+\epsilon \sin\tau)$.
We substitute this in the evolution equations and extract the first order terms as:
\begin{align}\label{eq:4s_lambda}
  \dot{\lambda}^{(1)}_1=-\frac{\mathcal{K}_1\mathbb{G}^{(0)}_1}{1+\mathcal{D}_1\mathbb{G}^{(0)}_1 }\lambda^{(1)}_1 +\frac{\mathbb{F}^{(0)}_1 \ell_2 \cos\tau}{1+\mathcal{D}_1\mathbb{G}^{(0)}_1}-\frac{\mathbb{H}^{(0)}_1}{1+\mathcal{D}_1\mathbb{G}^{(0)}_1}\dot{\lambda}^{(1)}_3, 
\nonumber
\\
  \dot{\lambda}^{(1)}_3=-\frac{\mathcal{K}_3\mathbb{G}^{(0)}_3}{1+\mathcal{D}_3\mathbb{G}^{(0)}_3 }\lambda^{(1)}_3 +\frac{\mathbb{F}^{(0)}_3 \ell_2 \cos\tau}{1+\mathcal{D}_3\mathbb{G}^{(0)}_3} -\frac{\mathbb{H}^{(0)}_3}{1+\mathcal{D}_3\mathbb{G}^{(0)}_3}\dot{\lambda}^{(1)}_1.
\end{align}
Here we focus on a symmetric natural configuration $\ell_1=\ell_3$ such that the corresponding zeroth-order geometric functions coincide, simplifying the structure of the coupled system. 
The solution of the coupled equations Eq.(\ref{eq:4s_lambda}) can be written as
\begin{align}\label{eq:4S_lambda-sol}
  \lambda^{(1)}_1= -A_1 \ell_2 \sin \tau - B_1 \ell_2 \cos\tau , 
\nonumber
\\
  \lambda^{(1)}_3= - A_3 \ell_2 \sin \tau - B_3 \ell_2 \cos\tau.
\end{align}
Here we have excluded the exponentially-decaying transient terms that vanish after a couple of deformation cycle.
The coefficients $A_i$ and $B_i$ (positive for all parameter values) correspond to the in-phase and out-of-phase components of the viscoelastic response, respectively, and are provided in the Appendix section.

\begin{figure}[t]
    \centering
        \includegraphics[width=0.9\linewidth]{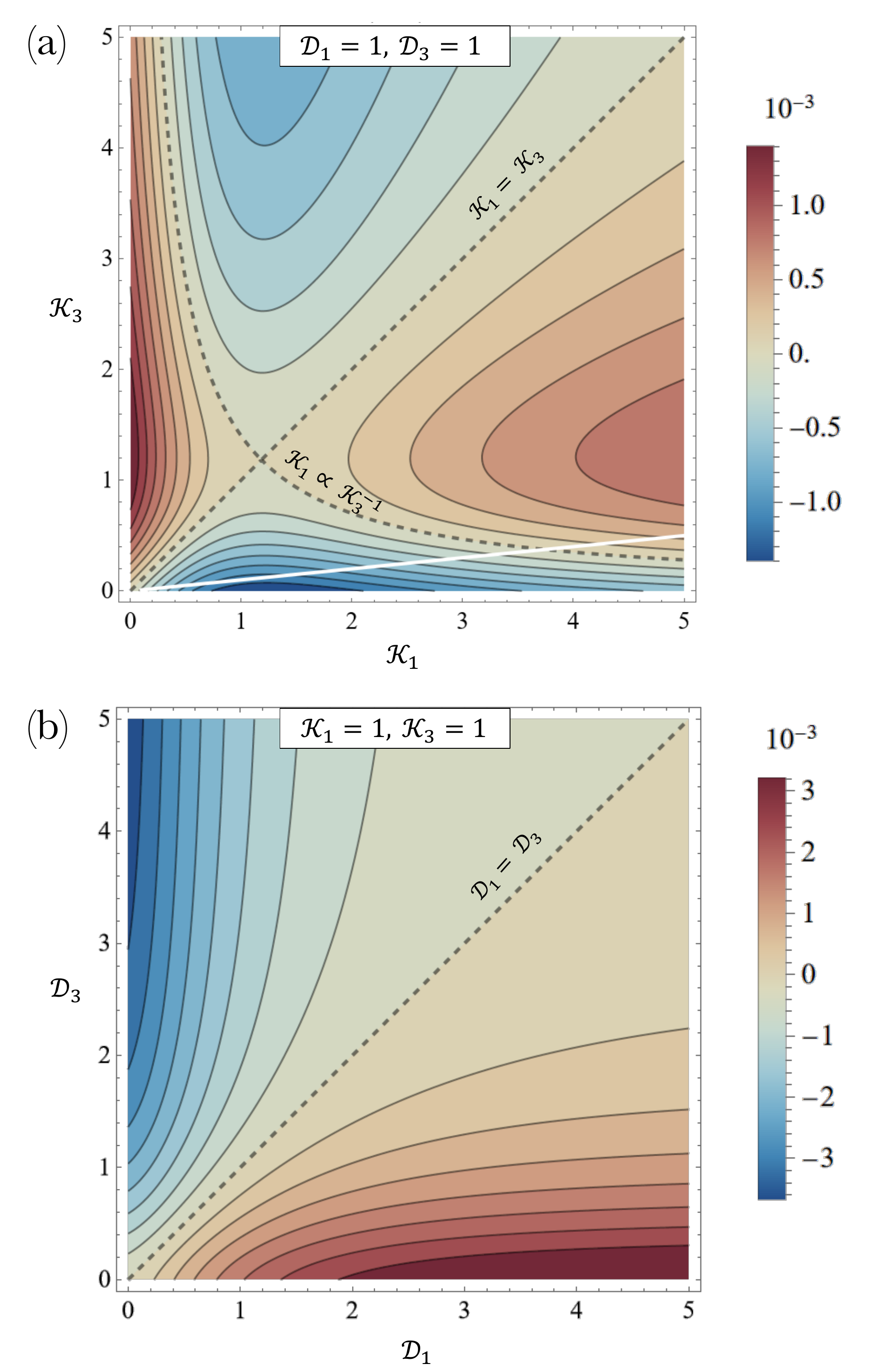}
    \caption{\footnotesize 
  Contour plots depicting variation of dimensionless displacement per unit cycle  with the two dimensionless elasticity numbers (a) and two damping numbers (b) of the passive viscoelastic links. The dashed lines depict zero contours. The white line depicts the connection with numerical results in Fig.\ref{fig:schematic1}(b).
  Parameters: $\ell_1=\ell_2=\ell_3=1/3$, $\alpha =0.1/6$.}
    \label{fig:schematic4}
\end{figure}

To obtain the leading-order expression for the mean velocity of a 4-sphere swimmer, we consider the dimensionless form derived by rearranging Eqs.~(\ref{eq:mob}-\ref{eq:kin}):
\begin{align}
    &\overline{v}=\frac{3 \alpha}{16} \left[
    \left(-\frac{\dot{\Lambda}_1+\dot{\Lambda}_3}{\Lambda_1+\Lambda_2}+\frac{\dot{\Lambda}_1+2\dot{\Lambda}_2}{\Lambda_3}+\frac{\dot{\Lambda}_1+\dot{\Lambda}_3}{\Lambda_2+\Lambda_3}+  \frac{\dot{\Lambda}_1-\dot{\Lambda}_3}{\Lambda_2}
    \right. \right. \nonumber
    \\
    &\left.\left.
   -\frac{2\dot{\Lambda}_2+\dot{\Lambda}_3}{\Lambda_1}+\frac{\dot{\Lambda}_3-\dot{\Lambda}_1}{\Lambda_1+\Lambda_2+\Lambda_3}\right) 
    + \left( 
    \frac{\dot{\Lambda}_3}{\Lambda_3} - \frac{\dot{\Lambda}_1}{\Lambda_1}
    \right)  \right]. 
\end{align}
Following a procedure analogous to that used to derive Eq.~(\ref{eq:3S_v}), we obtain the following leading order terms
\begin{align}\label{eq:4S_v}
    \overline{v}&=O(\epsilon)+\frac{3 \alpha \epsilon^2}{16}\left[\frac{(\lambda_1+\lambda_3+\ell_2\sin\tau)(\dot{\lambda}_1-\dot{\lambda}_3)}{(2\ell_1+\ell_2)^2}
    \nonumber
    \right.
    \\
    &+\left.\frac{(\dot{\lambda}_3-\dot{\lambda}_1)\sin\tau}{\ell_2}+ \frac{(\lambda_1-\lambda_3)(\dot{\lambda}_1+\dot{\lambda}_3)}{(\ell_1+\ell_2)^2}
    \right.
    \\
    &\left.
      +\frac{\lambda_1 (2 \ell_2\cos\tau+\dot{\lambda}_3)-\lambda_3 (2 \ell_2\cos\tau+\dot{\lambda}_1)}{\ell_1^2}
   \right].
\nonumber
\end{align}
Similar to the case of three-sphere swimmer, the $O(\epsilon)$ contribution does not lead to any net displacement upon time averaging.
Substituting Eq.(\ref{eq:4S_lambda-sol}) in Eq.(\ref{eq:4S_v}) and evaluating the time integral over one cycle yields
\begin{align}\label{eq:4S_X}
& x =\frac{3 \alpha \pi\epsilon^2}{16}\left[\left(\frac{2 \ell_2^2}{\ell_1^2}-\frac{\ell_2^2}{(2 \ell_1 + \ell_2)^2}+1\right)(B_3 - B_1)+
  \right.
    \\
    &\left.
    \frac{2 \ell_2^2
\left(7 \ell_1^4+14 \ell_2 \ell_1^3+13 \ell_2^2 \ell_1^2+6 \ell_2^3 \ell_1
+\ell_2^4\right)}{\ell_1^2 (\ell_1 + \ell_2)^2 (2 \ell_1 + \ell_2)^2
} (A_3 B_1 - A_1 B_3) \right].
\nonumber
\end{align}
The above expression shows that locomotion arises from out-of-phase deformations encoded through the coefficients $B_i$, as well as from their coupling with the in-phase responses. We find a close agreement of this expression with the numerical results for small deformations.
Fig.\ref{fig:schematic4} (a) shows the variation of displacement with the dimensionless elastic numbers, where two dashed lines depict the zeroes of net displacement obtained by solving $B_3-B_1=0$ and $A_3 B_1 - A_1 B_3=0$. 
At higher values of both $\mathcal{K}_1$ and $\mathcal{K}_3$ (implying lower actuation frequencies), corresponding to region above the hyperbolic dashed line, the resonance always occurs at the softer end. Consequently, the softer passive link undergoes larger  non-reciprocal deformations, thereby dictating the direction of locomotion. As a result, the swimmer displaces leftward (negative; blue region in Fig.\ref{fig:schematic4}a) toward link-1 when $\mathcal{K}_1 < \mathcal{K}_3$, the contrary applies for $\mathcal{K}_1 > \mathcal{K}_3$.
Conversely below the hyperbolic curve, at lower values of $\mathcal{K}_1$ and $\mathcal{K}_3$, the rapid actuation resonates with the stiffer ends; in this regime, the direction of locomotion shifts toward the link with the higher stiffness.
The white line relates the values of this contour with numerical results in Fig.\ref{fig:schematic2} (b).
Furthermore, the hyperbolic curve separating the regions grows as damping coefficient increases.
Finally, consistent with the numerical results discussed earlier, Fig.\ref{fig:schematic4} (b) depicts that side with lower damping coefficient will direct the locomotion across entire parameter space.

\section{Extracting far-field signature}\label{sec:4}
In addition to producing net propulsion, cyclic deformations of the swimmer generate a time-averaged flow field in the ambient fluid. 
The characteristics of the flow field define the swimmer type (e.g., pusher, puller, or neutral), which governs the inter-swimmer and swimmer-confinement hydrodynamic interactions.
To determine the resulting flow field, we construct the instantaneous velocity field as a superposition of the flow fields generated by the individual spheres. We recognize that Eq.(\ref{eq:mob}) can be re-written in terms of the Oseen tensor $\mathcal{G}_{ij}$. Using the non-dimensional scaling consistent with \S\ref{sec:3}, we write
\begin{align}\label{eq:vel-oseen}
    v_i(\IB{r},\tau) =& \sum_{a=1}^{n} \frac{\mathcal{G}_{ij}[\IB{r}-\IB{r}_a(t)]  f_{a} (\tau) \hat{p}_j}{8 \pi}, \; \text{where}
    \\
\mathcal{G}_{ij} &=\frac{1}{|\IB{r}-\IB{r}_a(\tau)|}\left(\delta_{ij}+\hat{r}_i \hat{r}_j\right).
\nonumber
\end{align}
Here, $f_{a}=|\IB{f}_a|$ denotes the force monopole due to the motion of sphere $a$ along the propulsion axis $\hat{\IB{p}}$, the position vector is written as $\hat{\IB{r}}$, and $\IB{r}_a=r_a(\tau)\,\hat{\IB{p}}$ such that $r_{a+1}-r_{a}= \Lambda_a$.
 Next, we write the time-averaged flow field by integrating the instantaneous flow field over one swimming period,
\begin{equation}\label{eq:vel-time_avg}
\overline{v}_i(\IB{r})=\frac{1}{2\pi}\int_0^{2\pi} v_i(\IB{r},\tau)\,\mathrm{d}\tau.
\end{equation}
However, analytical evaluation of this integral is challenging because the positions of the spheres, $r_a$, evolve in time, rendering the Oseen tensor explicitly time dependent. As a result, direct integration will not yield tractable analytical expressions.
To extract useful insights, we employ a multipole expansion of the swimmer-induced flow field about the observation point $r$, which allows us to characterize the far-field behavior of the swimmer. We choose the initial position of the second sphere to be at the origin.
Under the far-field limit, the distance to the observation point is assumed to be much larger than the displacements of spheres and the swimmer over one cycle of deformation.
Under this assumption, the Oseen tensor for each sphere may be expanded as a series of $r_a$,
\begin{equation}
\mathcal{G}_{ij}[\IB{r}-\IB{r}_a(\tau)]=\sum_{m=0}^{\infty}\frac{(-1)^m}{m!}\left(\mathcal{G}_{ij,k_1...k_m}\,\hat{p}_{k_1}...\hat{p}_{k_m}\right)r_a(\tau)^{m} \nonumber
\end{equation}
Here, the $m^{\text{th}}$ order outer product in $\IB{r}_a$ is simplified as: $\IB{r}_{a} \IB{r}_a \cdots \IB{r}_{a} = \left( \hat{p}_{k_1} \hat{p}_{k_2} \cdots \hat{p}_{k_m} \right) {r_a}^m $.
Substituting this into Eqs. (\ref{eq:vel-oseen}) and (\ref{eq:vel-time_avg}) yields the multipole expansion of the time-averaged flow field as
\begin{align}
\overline{v}_i({r})&=\sum_{m=0}^{\infty}\frac{(-1)^m}{m!}\,\mathcal{G}_{ij,k_1\cdots k_m}\,\hat{p}_{k_1}\cdots \hat{p}_{k_m} 
\nonumber
\\
&\quad
\times \left[\frac{1}{16 \pi^2}\int_0^{2\pi}\sum_{a=1}^{n}f_{a}(\tau) \hat{p}_j\, r_a^{m}(\tau)\,\mathrm{d}t\right]
\end{align}
Here, the first term describes the spatial decay of the flow field in $m$ multipoles, whereas the second bracketed term contains the moments of the forces and provides the time-averaged strength of the corresponding multipole.
The monopole term ($m=0$) vanishes due to the force-free condition. 
Consequently, the far-field description can be fairly represented by the dipolar ($m=1$) field\footnote{Higher order quadrupolar ($m=2$) contributions are provided in the Appendix section.}:
\begin{align}\label{eq:vel-multipole-main}
\overline{v}_i({r})
&=-\mathcal{G}_{ij,k_1} \hat{p}_{k_1} \hat{p}_j\left[
\frac{1}{16 \pi^2}\int_0^{2\pi}\sum_{a=1}^{n}
f_{a}(\tau) r_a(\tau)\mathrm{d}\tau\right]
\\
&
+\frac{1}{2}\,\mathcal{G}_{ij,k_1 k_2} \hat{p}_{k_1} \hat{p}_{k_2}\hat{p}_j\left[\frac{1}{16 \pi^2}
\int_0^{2\pi}\sum_{a=1}^{n}f_{a}(\tau) r_a^{2}(\tau)\mathrm{d}\tau
\right]
\nonumber
\end{align}

\begin{figure}[t]
    \centering
        \includegraphics[width=0.8\linewidth]{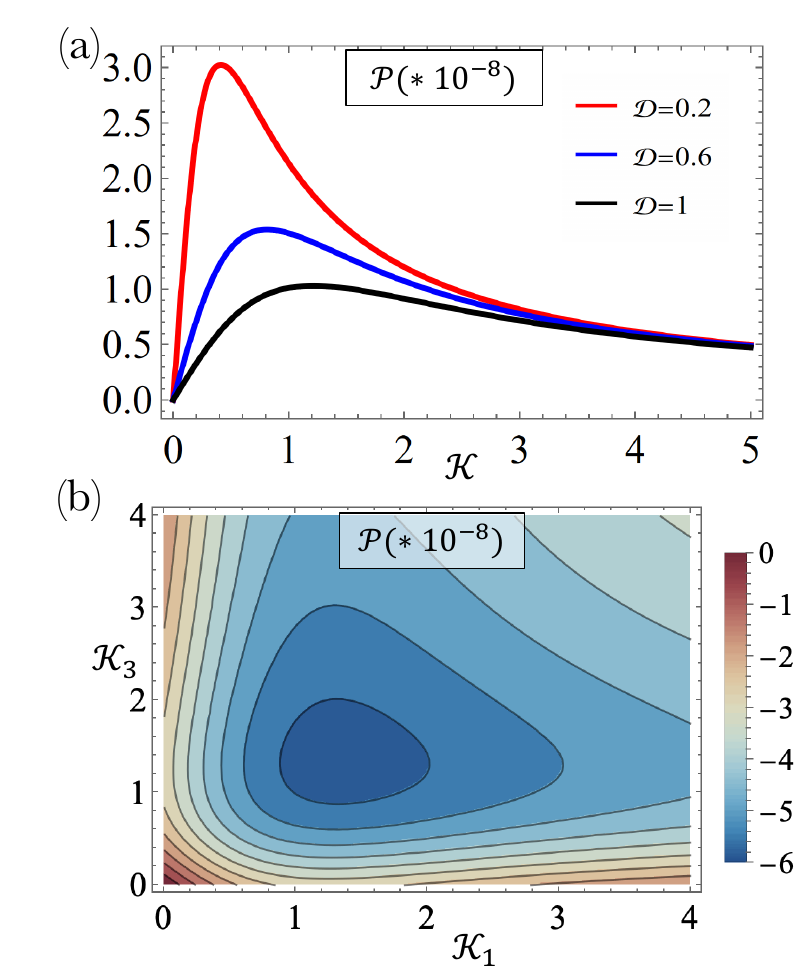}
    \caption{\footnotesize 
    (a) Dipole Strength variation with elasticity numbers ($\mathcal{K} $) for three different damping coefficients ($\mathcal{D} $) with geometrical values: $\ell_1=2\ell_2=2/3$ and $\alpha=0.1/6$
    (b) Contour plot depicting variation of Dipole strength with the two dimensionless elasticity numbers with following parameters: $\mathcal{D}_1 =\mathcal{D}_3 = 1$, $\ell_1 = \ell_2 = \ell_3 = 1/3$ and $\alpha = 0.1/6$.
  }
    \label{fig:schematic5}
\end{figure}

The force dipole signature is given by:
\begin{equation}
\mathcal{G}_{ij,k_1} \hat{p}_{k_1} \hat{p}_j = \frac{1}{r^2}\left[1 -  3(\hat{\IB{p}}.\hat{\IB{r}})^2 \right]\hat{r}_i   
\end{equation}
We next substitute the results from \S\ref{sec:3} in Eq. (\ref{eq:vel-multipole-main}).
For the 3-sphere swimmer ($N=3$), we evaluate the strength of dipole $\mathcal{P}$ as
\begin{align}
\mathcal{P}&=  \frac{1}{16 \pi^2}\int_0^{2\pi} \left[ f_1 r_1+ f_2 r_2+ f_3 r_3\right]\mathrm{d}\tau
\nonumber
\\
 &=  \frac{1}{16 \pi^2}  \int_0^{2\pi}  \left[ -f_1 \Lambda_1+ f_3 \Lambda_2\right]\mathrm{d}\tau ,
\end{align}
where we have used the force-free criteria to substitute $f_2= - f_1 - f_3$.
Substituting the expressions for $f_1$, $f_3$, $\Lambda_1$, and $\Lambda_2$ into the above equation and carrying out the integration over one period, we retain only the leading-order term at $O(\epsilon^2)$.
\begin{equation}\label{eq:dipole}
\mathcal{P}
=
\frac{
\epsilon^{2} \alpha^{2} B 
\left(2 \ell_1^{4}+ 7 \ell_2 \ell_1^{3}+ 11 \ell_2^{2} \ell_1^{2}+ 7 \ell_2^{3} \ell_1+ 2 \ell_2^{4}\right) (\ell_1 - \ell_2) }{
16\, \ell_1^{2} (\ell_1 + \ell_2)^{2}}
\end{equation}
where $O(\epsilon^2 \alpha^3)$ and higher order terms contribute negligibly.
We note that the factor responsible for out-of-phase deformation, and thus the locomotion ($B$), also governs the magnitude of dipolar strength.
Furthermore, since $B$ and the other factors are always positive, the sign of dipole is determined by the geometric anisotropy ($\ell_1 - \ell_2$).
In Fig. \ref{fig:schematic5}(a), this relationship between dipole strength and elasticity number follows a trend expected of factor $B$: because non-reciprocity vanishes during both quasi-static ($\mathcal{K} \to \infty$) and high-speed ($\mathcal{K} \to 0$) actuations, the dipolar signature effectively disappears at these extremes.

In a similar manner, for the 4-sphere swimmer that exhibits symmetry in the natural configuration, we obtain
\begin{equation}\label{eq:quadrupole}
\mathcal{P}
=
\frac{-9 \epsilon^{2} \alpha^{2}\left( B_1 + B_3 \right) \ell_2^{3}\left(3 \ell_1^{2}
+ 3 \ell_2 \ell_1+ \ell_2^{2}\right)}{32\, \ell_1^{2} (\ell_1 + \ell_2)^{2}},
\end{equation}
which denotes a puller for all parameter values, irrespective of direction of locomotion.
Fig. \ref{fig:schematic5}(b) shows that, similar to 3-sphere case, there exists an optimal frequency that maximizes the out-of-phase factor $B_1+B_3$.


\section{Conclusion}
In this study, we have characterized the influence of viscoelasticity on the kinematics and far-field signature of linked-sphere microswimmers.
For the three-sphere swimmer, viscoelastic deformability of the passive link introduces a frequency-dependent phase lag that governs net locomotion, with optimal displacement arising from the trade-off between deformation magnitude and non-reciprocal phase asymmetry.
Asymptotic expansion for weak $O(\epsilon)$  deformation regime reveals that leading order locomotion is $O(\epsilon^2 \alpha)$ and is shown to be directly proportional to the degree of phase asymmetry between the passive and active links ($B$ in Eq. \ref{eq:3S_X}). Here $\alpha$ is ratio of sphere radius to swimmer's natural length.
For the four-sphere swimmer, asymmetry in viscoelastic or geometric properties between the two passive links is a necessary condition for locomotion, and gives rise to a critical frequency at which the direction of swimming reverses.
The asymptotic expression (Eq. \ref{eq:4S_X}) for net displacement confirms that locomotion arises from out-of-phase deformations and their coupling with in-phase responses. 
The contour plots of displacement further delineate the parameter space, revealing four distinct regions demarcated by the intersection of hyperbolic and linear zero-contour boundaries in the space of viscoelastic properties of the two passive links.
Finally, within the asymptotic framework, we employ multipole expansion to derive closed-form expressions for the $O(\epsilon^2 \alpha^2)$ far-field dipolar and quadrupolar hydrodynamic signatures (Eqs.~\ref{eq:dipole},\ref{eq:quadrupole}), characterizing the swimmer type and its influence on the surrounding fluid.
These results collectively demonstrate that both the magnitude and direction of locomotion, as well as the far-field hydrodynamic signature, of viscoelastic microswimmers can be systematically programmed through the material properties and geometry of the passive links, offering a rich and tunable parameter space for the design of envisioned microbots.
\\

\noindent
\textbf{Acknowledgments}.
The authors thank the Indian Institute of Technology Kanpur for support
via Initiation Grant (IITK-CHE-2023066). 
\\\\
\textbf{Declarations}. The authors report no conflict of interest.
All data supporting the findings of this study are contained within the manuscript. Numerical integration and evaluation of the analytical expressions were performed using Wolfram Mathematica.

\appendix

\section{Coefficients for 4-sphere swimmer}

{
\small
\begin{widetext}
\begin{equation}
\mathbb{H}^{(0)}_1
=\frac{9 \alpha ^2 \left(-2 \ell_1^5+\ell_2 \ell_1^4+8 \ell_2^2 \ell_1^3+7 \ell_2^3 \ell_1^2+4 \ell_2^4\ell_1+\ell_2^5\right)-12 \alpha  \ell_1 \ell_2^2 (\ell_1+\ell_2)\left(3 \ell_1^2+3 \ell_2 \ell_1+\ell_2^2\right)+4 \ell_1^2 \ell_2^2 (\ell_1+\ell_2)^2 (2 \ell_1+\ell_2)}{\mathbb{D}_1
}
\end{equation}
\begin{equation}
\mathbb{F}^{(0)}_1
=\frac{2}{\mathbb{D}_1} \left(\ell_1^3 (3 \alpha +4 \ell_2)+6 \ell_2 \ell_1^2 (\ell_2-\alpha )+\ell_2^2 \ell_1 (2 \ell_2-9 \alpha )-3 \alpha  \ell_2^3\right)
\left(3 \alpha  (\ell_1^2+\ell_2 \ell_1+\ell_2^2)-2 \ell_1 \ell_2(\ell_1+\ell_2)\right)
\end{equation}
\begin{align}
\mathbb{G}^{(0)}_1
&=\frac{\left(\ell_1^3 (3 \alpha +4 \ell_2)+6 \ell_2 \ell_1^2 (\ell_2-\alpha )+\ell_2^2 \ell_1 (2 \ell_2-9 \alpha )-3 \alpha  \ell_2^3\right)}
{3 \pi  \alpha  \ell_1 (\ell_1+\ell_2) (2 \ell_1+\ell_2) \mathbb{D}_1} \left[9 \alpha ^2 (\ell_1^4+2 \ell_2 \ell_1^3+\ell_2^2 \ell_1^2-2 \ell_2^3 \ell_1-\ell_2^4) \right.
\nonumber
\\
& \qquad \qquad \qquad \qquad \qquad\qquad \qquad \qquad \qquad\qquad \qquad \qquad \qquad \left. -12 \alpha  \ell_1^2 (\ell_1+\ell_2)^3+4 \ell_1^2 \ell_2 (2 \ell_1+\ell_2) (\ell_1+\ell_2)^2\right]
\end{align}
\begin{equation}
B_1=\frac{-\mathbb{F}^{(0)}_1 \mathbb{G}^{(0)}_1}{\mathbb{D}_2} \Big[- \mathbb{H}^{(0)}_1 \mathcal{K}_1 \left( \mathbb{G}^{(0)}_1 \mathcal{D}_3 + 1 \right)+ \mathbb{H}^{(0)}_1 \mathcal{K}_3 \left( - \mathbb{G}^{(0)}_1 \mathcal{D}_1 + \mathbb{H}^{(0)}_1 - 1 \right)+ \mathbb{G}^{(0)}_1 \mathcal{K}_1 \left( \mathbb{G}^{(0)}_1 \left( \mathcal{K}_3^2 + \mathcal{D}_3^2 \right) + 2 \mathcal{D}_3 \right)+ \mathcal{K}_1\Big]
\end{equation}
\begin{align}
A_1 &=\frac{-\mathbb{F}^{(0)}_1}{\mathbb{D}_2} \left[\mathbb{H}^{(0)}_1 \Big(\mathbb{G}^{(0)\,2}_1 \left( \mathcal{K}_1 \mathcal{K}_3 - \mathcal{D}_1 \mathcal{D}_3 \right)- \mathbb{G}^{(0)}_1 \left( \mathcal{D}_1 + \mathcal{D}_3 \right)- 1\Big) - \mathbb{H}^{(0)\,2}_1 \left( \mathbb{G}_1 \mathcal{D}_3 + 1 \right) \right. \nonumber
\\
&
\qquad \qquad \qquad   
\left.
+ \left( \mathbb{G}^{(0)}_1 \mathcal{D}_1 + 1 \right)  \left(  \mathbb{G}^{(0)\,2}_1 \left( \mathcal{K}_3^2 + \mathcal{D}_3^2 \right)  + 2 \mathbb{G}^{(0)}_1 \mathcal{D}_3  + 1  \right)+ \mathbb{H}^{(0)\,3}_1\right]
\\
B_3 &=\frac{-\mathbb{F}^{(0)}_1 \mathbb{G}^{(0)}_1}{\mathbb{D}_2} \Big[- \mathbb{H}^{(0)}_1 \left(\mathbb{G}^{(0)}_1 \mathcal{K}_1 \mathcal{D}_3
+ \mathbb{G}^{(0)}_1 \mathcal{K}_3 \mathcal{D}_1+ \mathcal{K}_1+ \mathcal{K}_3\right)+ \mathbb{G}^{(0)}_1 \mathcal{K}_3 \left(\mathbb{G}^{(0)}_1 \left( \mathcal{K}_1^2 + \mathcal{D}_1^2 \right)+ 2 \mathcal{D}_1\right)+ \mathbb{H}^{(0)\,2}_1 \mathcal{K}_1+ \mathcal{K}_3\Big]
\\
A_3 &=\frac{-\mathbb{F}^{(0)}_1}{\mathbb{D}_2} \left[\mathbb{H}^{(0)}_1 \Big(\mathbb{G}^{(0)\,2}_1 \left( \mathcal{K}_1 \mathcal{K}_3 - \mathcal{D}_1 \mathcal{D}_3 \right)- \mathbb{G}^{(0)}_1 \left( \mathcal{D}_1 + \mathcal{D}_3 \right)- 1\Big)- \mathbb{H}^{(0)\,2}_1 \left( \mathbb{G}^{(0)}_1 \mathcal{D}_1 + 1 \right) \right. \nonumber
\\
& \qquad\qquad\qquad  \left. + \left( \mathbb{G}^{(0)}_1 \mathcal{D}_3 + 1 \right)  \left(  \mathbb{G}^{(0)\,2}_1 \left( \mathcal{K}_1^2 + \mathcal{D}_1^2 \right)  + 2 \mathbb{G}^{(0)}_1 \mathcal{D}_1  + 1  \right)+ \mathbb{H}^{(0)\,3}_1\right],
\end{align}
\\
where
\begin{align}
     \mathbb{D}_1 &= 3 (2 \ell_1+\ell_2)\left[-3 \alpha ^2 \left(\ell_1^4-2 \ell_2 \ell_1^3-5 \ell_2^2 \ell_1^2-2 \ell_2^3 \ell_1+\ell_2^4\right)-4 \alpha  \ell_1 \ell_2 (\ell_1+\ell_2)\left(\ell_1^2+3 \ell_2 \ell_1+\ell_2^2\right)+4 \ell_1^2 \ell_2^2 (\ell_1+\ell_2)^2\right]
     \nonumber
     \\
     \mathbb{D}_2 &= - 2 \mathbb{H}^{(0)\,2}_1\Big[\mathbb{G}^{(0)}_1 \left(- \mathbb{G}^{(0)}_1 \mathcal{K}_1 \mathcal{K}_3+ \mathbb{G}^{(0)}_1 \mathcal{D}_1 \mathcal{D}_3+ \mathcal{D}_1 + \mathcal{D}_3\right) + 1\Big] \nonumber
     \\
& \qquad + \left(\mathbb{G}_1^{(0)2}\left( \mathcal{K}_1^2 + \mathcal{D}_1^2 \right)+ 2 \mathbb{G}_1^{(0)}\mathcal{D}_1 + 1\right)*\left(\mathbb{G}_1^{(0)2}\left( \mathcal{K}_3^2 + \mathcal{D}_3^2 \right)+ 2 \mathbb{G}_1^{(0)} \mathcal{D}_3 + 1\right)
+ \mathbb{H}^{(0)\,4}_1.
\nonumber
\end{align}

\end{widetext}
}

\begin{figure}[t]
    \centering
    \includegraphics[width=0.7\linewidth]{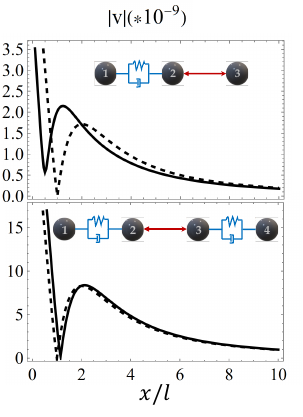}
    \caption{\footnotesize
    Comparison between numerical simulation(solid line) and analytical solution(dashed line) for the magnitude of velocity field over \textit{x}-coordinate for $y=z=l$. (top) Parameters: $\mathcal{K}=\mathcal{D}=1$,$\ell_1=2\ell_2=1/3$ and $\alpha=0.1/6$
    (bottom)  Parameters: $\mathcal{K}_1=2\mathcal{K}_3=1$,$\mathcal{D}_1=\mathcal{D}_3=1$,$\ell_1=\ell_2=\ell_3=1/3$ and $\alpha=0.1/6$
    }
    \label{fig:appendix-fig}
\end{figure}

\section{Quadrupolar flow signature}
The $O(1/r^3)$ quadrupolar field is given by
\begin{align}
     &\frac{1}{2}\mathcal{G}_{ij,k_1 k_2}\,\hat{p}_{k_1} \hat{p}_{k_2}\hat{p}_j =
     \nonumber
     \\
&\quad
     \frac{1}{2 r^3}\left[3(\hat{\IB{p}}.\hat{\IB{r}})(5(\hat{\IB{p}}.\hat{\IB{r}})^2-3)\hat{r}_i-(3(\hat{\IB{p}}.\hat{\IB{r}})^2-1)\hat{p}_i\right]
\end{align}
We next substitute the results from \S\ref{sec:3} in Eq. (\ref{eq:vel-multipole-main}).
For the 3-sphere swimmer ($N=3$), we evaluate the strength of quadrupole $\mathcal{Q}$ as
\begin{align}
    \mathcal{Q}&=\frac{1}{16 \pi^2}
\int_0^{2\pi}\left[f_1 r_1^2+ f_2 r_2^2+ f_3 r_3^2\right]\mathrm{d}\tau
\\
&= \frac{1}{16 \pi^2}
\int_0^{2\pi}\left[-f_1(2 r_2-\Lambda_1)\Lambda_1+f_3(2 r_2+\Lambda_2)\Lambda_2\right]\mathrm{d}\tau.
\nonumber
\end{align}
In above, we expressed $r_1$ and $r_3$ in terms of $\Lambda_1$, $\Lambda_2$ and $r_2$; the position $r_2$ is obtained by integrating $v_2$ over the interval $[0,\tau]$.
Substituting the expressions for $f_1$, $f_3$, $\Lambda_1$,$\Lambda_2$ and $r_2$  into the above equation and carrying out the integration over one period, we retain only the leading $O(\epsilon^2 \alpha^2)$ term.
\begin{equation}
\mathcal{Q}=\frac{- \epsilon^{2} \alpha^{2} B 
\left(2 \ell_1^{4}+ 4 \ell_2 \ell_1^{3}+ 5 \ell_2^{2} \ell_1^{2}+ 4 \ell_2^{3} \ell_1+ 2 \ell_2^{4}\right) \left(\ell_1^{2} + \ell_2^{2}\right) }{
16 \ell_1^{2} (\ell_1 + \ell_2)^{2}}
\end{equation}
For a 4-sphere swimmer,
\begin{align}
 \mathcal{Q}&=\frac{1}{16 \pi^2}
\int_0^{2\pi}\left[f_1 r_1^2+ f_2 r_2^2+ f_3 r_3^2+f_4r_4^2\right]\mathrm{d}\tau
\nonumber
\\&= \frac{1}{16 \pi^2}
\int_0^{2\pi}\left[-f_1(2 r_2-\Lambda_1)\Lambda_1+f_3(2 r_2+\Lambda_2)\Lambda_2
\nonumber
 \right.
    \\
    &\left.
+f_4(2r_2+\Lambda_2+\Lambda_3)(\Lambda_2+\Lambda_3)\right]\mathrm{d}\tau
\end{align}
Substituting the expressions for $f_1$, $f_3$, $f_4$, $\Lambda_1$, $\Lambda_2$, $\Lambda_3$ and $r_2$ into the above equation and carrying out the integration over one period, we retain only the leading-order term.
\begin{align}
\mathcal{Q}&=\frac{9 \epsilon^2 \alpha^2  (A_3 B_1 - A_1 B_3) \left(9 \ell_1^2 + 9 \ell_1 \ell_2 + 2 \ell_2^2 \right) \ell_2^2 }
{32 (2 \ell_1 + \ell_2)^2} 
 \nonumber
\\
&- \frac{9 \alpha^2\, \ell_2^7\, \epsilon^2 (B_1 + B_3)(7 \ell_1 + \ell_2)}
{32\, \ell_1^2 (\ell_1 + \ell_2)^2 (2 \ell_1 + \ell_2)^2}
 \nonumber
\\
&- \frac{9 \alpha^2\, \ell_1^3\, \epsilon^2 (B_1 - B_3)\left(2 \ell_1^3 + 8 \ell_1^2 \ell_2 + 21 \ell_1 \ell_2^2 + 35 \ell_2^3 \right)}
{32 (\ell_1 + \ell_2)^2 (2 \ell_1 + \ell_2)^2}
 \nonumber
\\
&- \frac{9 \alpha^2\, B_1\, \ell_2^4\, \epsilon^2 \left(43 \ell_1^2 + 37 \ell_1 \ell_2 + 21 \ell_2^2 \right)}
{32 (\ell_1 + \ell_2)^2 (2 \ell_1 + \ell_2)^2}
 \nonumber
\\
&- \frac{9 \alpha^2\, B_3\, \ell_2^4\, \epsilon^2 \left(-19 \ell_1^2 + 11 \ell_1 \ell_2 + 17 \ell_2^2 \right)}
{32 (\ell_1 + \ell_2)^2 (2 \ell_1 + \ell_2)^2}
\end{align}
For both cases of swimmers, Fig.~\ref{fig:appendix-fig} shows how the far-field signatures (combination of $\mathcal{P}$ and $\mathcal{Q}$ contributions) match the numerical flow fields beyond swimmer length scale.

\bibliography{references}

\end{document}